\documentclass[twocolumn,showpacs,showkeys,preprintnumbers,amsmath,amssymb]{revtex4-1}
\usepackage{bbm}
\usepackage{amsfonts}

\usepackage{epsfig}
\usepackage{graphicx}
\usepackage{amssymb}   
\usepackage{dcolumn}
\usepackage{bm}
\usepackage[colorlinks,citecolor=blue, linkcolor=red,hyperindex]{hyperref}
\begin{document}

\vspace{2.5cm}
\title{Improving the precision of multiparameter estimation in the teleportation of qutrit under amplitude damping noise}
\author{Yan-Ling Li$^{1}$}
\altaffiliation{liyanling0423@gmail.com}
\author{Yi-Bo Zeng$^{1}$}
\author{Lin Yao$^{1}$}
\author{Xing Xiao$^{2}$}
\altaffiliation{xiaoxing@gnnu.edu.cn}
\affiliation{$^{1}$ School of Information Engineering, Jiangxi University of Science and Technology, Ganzhou, China\\
$^{2}$ College of Physics and Electronic Information, Gannan Normal University, Ganzhou, China}

\begin{abstract}
Since the initial discovery of quantum teleportation, it is devoted to transferring unknown quantum states from one party to another distant partner. However, in the scenarios of remote sensing, what people truly care about is the information carried by certain parameters. The problem of multiparameter estimation in the framework of qutrit teleportation under amplitude damping (AD) noise is studied. Particularly, two schemes are proposed to battle against AD noise and enhance the precision of multiparameter estimation by utilizing weak measurement (WM) and environment-assisted measurement (EAM). For two-phase parameters encoded in a qutrit state, the analytical formulas of the quantum Fisher information matrix (QFIM) can be obtained. The results prove that the scheme of EAM outperforms the WM one in the improvements of both independent and simultaneous estimation precision. Remarkably, the EAM scheme can completely ensure the estimation precision against the contamination by AD noise. The reason should be attributed to the fact that EAM is carried out after the AD noise. Thus, it extracts information from both the system and the environment. The findings show that the techniques of WM and EAM are helpful for remote quantum sensing and can be generalized to other qutrit-based quantum information tasks under AD decoherence.
\end{abstract}
\keywords{quantum teleportation, multiparameter estimation, weak measurement, environment-assisted measurement}
\maketitle

\section{Introduction}
\label{sec1}
Quantum metrology devotes to seeking scenarios where quantum resources (entanglement, squeezing and so on) can provide enhancements in the parameter estimation over the classical strategies.\textsuperscript{\cite{Giovannetti2011,Taylor2016,Polino2020}} The ultimate precision of single parameter is determined by the quantum Cram\'er-Rao bound (CRB).\textsuperscript{\cite{Braunstein1994}} It is well known that single parameter quantum CRB can be always saturated by optimizing over all valid quantum measurements. However, most high level applications intrinsically involve multiple unknown parameters, such as quantum imaging, detection of classical electric field, magnetic field and gravitational field. All of them fall in the subject of multiparameter estimation. The most important reason for the general interest of multiparameter estimation is that simultaneous quantum estimation of multiple phases can provide better precision than estimating them individually.\textsuperscript{\cite{Szczykulska2016,Gessner2018,Liu2019}} The quantum multiparameter CRB characterizes the precision of simultaneous estimation. The saturation of multiparameter CRB is an open question because different parameters usually have different optimal measurements which may not commute with each other.

Except for the saturability problem, it also should be noted that the promise of quantum advantages in parameter estimation is limited by the presence of noise in any realistic experiment, such as dissipation and dephasing.\textsuperscript{\cite{Escher2011a,Escher2011b}} Particularly, dissipation plays a crucial role in experiments involving photon loss and atomic decay. The results confirm that
it is definitely a reduction of the precision when the noise is considered, but the quantum parameter estimation is still possible to surpass the classical schemes. In more recent developments, the question of how to deal with the noise in quantum parameter estimation was addressed. \textsuperscript{\cite{Demkowicz2012,Chin2012,Tsang2013,Alipour2014,Hu2020a}} Various approaches for protecting parameter precision against noise have been proposed, such as quantum error
correction,\textsuperscript{\cite{Arrad2014,Dur2014,Kessler2014}} dynamical decoupling,\textsuperscript{\cite{Tan2013}} optimal feedback control,\textsuperscript{\cite{Liu2017}} correlated effects of the noisy channels\textsuperscript{\cite{Jin2018,Hu2020b}} and non-Markovianity of the environments.\textsuperscript{\cite{Bellomo2007,Li2015,Xu2022}}

In addition to the strategies mentioned above, weak
measurement (WM)\textsuperscript{\cite{Kim2012,Man2012,Li2013,Wang2014a,Li2016,Xiao2016a,Xiao2016b,Li2017,Xiao2018,Xiao2020,Im2021,Li2022}} and environment-assited measurement (EAM)\textsuperscript{\cite{Zhao2013,Wang2014b,Xu2015,Li2019,Li2021}} are receiving increasing attentions as new techniques to combat the dissipative noise. A WM operation is an extension of the traditional von Neumann projective measurement. It is reversible because the measured system does not totally collapse to the eigenstate of the measurement operator. Thus, the initial quantum state could be recovered by post quantum measurement reversal (QMR) operation with a certain probability. This merit enables WM to be one of the potential candidates for further suppression of the noise in quantum information processing. Similarly, the combination of EAM and QMR also devotes to recovering the initial quantum states. So far, the power of WM and EAM has been demonstrated
in protecting entanglement,\textsuperscript{\cite{Kim2012,Man2012,Xiao2016a}} quantum discord\textsuperscript{\cite{Li2013,Xiao2018,Jebli2020}} and other quantum correlations.\textsuperscript{\cite{Hu2018}} Particularly, the proposals for improving the precision of single-parameter quantum estimation by the WM and EAM have been proposed in recent studies.\textsuperscript{\cite{Xiao2016b,Li2021,Jin2019}} For multiparameter quantum estimation, systematic study of WM and EAM to enhance the precision under decoherence is still lacking.

Although the problem of multiparameter quantum estimation could be illustrated in the single-qubit system in proof of principle (e.g., the simultaneous estimation of weight and phase parameters), it is more naturally to estimate multiple phase parameters in higher dimensional systems. Qutrit-based (and, more generally, qudit-based) systems have been confirmed to provide significant advantages in the context of quantum technology, such as quantum communications,\textsuperscript{\cite{Cerf2002,Ali2007,Cozzolino2019}} quantum error correction,\textsuperscript{\cite{Campbell2014}} quantum simulation,\textsuperscript{\cite{Neeley2009}} quantum computation\textsuperscript{\cite{Lanyon2009}} and
high-fidelity magic state distillation.\textsuperscript{\cite{Campbell2012}} Moreover, it is of particular interest to note that the qutrit-based system can offer an enhanced precision of magnetic-field measurement.\textsuperscript{\cite{Shlyakhov2018}}

Motivated by the above considerations, we investigate the problem of multiple phases estimation in the framework of qutrit teleportation under the influence of amplitude damping (AD) noise. We focus on how to improve the estimation precision in the qutrit teleportation with the assistance of WM and EAM. The explicit expression of the quantum Fisher information matrix (QFIM) of phase parameters $\phi_{1}$ and $\phi_{2}$ encoded in the teleported qutrit is obtained. We shall show that both WM and EAM can improve the precisions of individual and simultaneous estimations in a probatilistic way. The performance of simultaneous estimation is better than individual estimation under either WM scheme or EAM scheme. Finally, the efficiency analysis proves that the EAM scheme outperforms the WM one. Our results provide two possible strategies for suppressing the AD decoherence and enhancing the precision of multiple phases estimation using qutrit-based quantum states, in which the multiparameter nature of the problem leads to an intrinsic benefit when exploiting the techniques of WM and EAM.

This paper is organized as follows: In Section \ref{sec2},  we briefly review the theories of quantum multiparameter estimation, AD noise, WM, QMR, and quantum gates for qutrit. In Section \ref{sec3}, we present the teleportation of an unknown qutrit in the presence of AD noise and discuss the estimation precision of phase parameters encoded in the teleported qutrit. In order to suppress the influence of AD noise, we propose a scheme to improve the precision of phase parameters estimation by WM and QMR in Section \ref{sec4}. Moreover, in Section \ref{sec5}, we exploit another improved scheme drawing support from EAM and QMR. Finally, a comprehensively comparison of the two schemes from the perspective of multiparameter estimation and a brief summary are given in Section \ref{sec6}.
\section{Preliminaries}
\label{sec2}

\subsection{Quantum multiparameter estimation}
Assuming that the general case of simultaneously estimating parameters $\boldsymbol{\theta}=(\theta_{1}, \ldots, \theta_{a}, \ldots, \theta_{b}, \ldots,\theta_{n})^{\mathrm{T}}$ are encoded in a $d$-dimensional density matrix $\rho=\rho(\boldsymbol{\theta})$. Further, let $\boldsymbol{\hat{\theta}}=(\hat{\theta}_{1}, \ldots, \hat{\theta}_{a}, \ldots, \hat{\theta}_{b}, \ldots,\hat{\theta}_{n})^{\mathrm{T}}$ be an estimator of $\theta$, and $\hat{\theta}_{a}$ is the estimator of $\theta_{a}$. For unbiased estimators, the quantum CRB establishes a lower bound to the covariance matrix in terms of
the QFIM.\textsuperscript{\cite{Braunstein1994,Helstrom1969}}
\begin{eqnarray}
\label{eq1}
\begin{split}
\operatorname{Cov}(\boldsymbol{\hat{\theta}}) \geq \mathcal{F}^{-1}(\boldsymbol{\theta})
\end{split}
\end{eqnarray}
where $\mathcal{F}(\boldsymbol{\theta})$ is the QFIM. The entries of QFIM for the parameters $\theta_{a}$ and $\theta_{b}$ with the spectral decomposition $\rho=\sum_{i=0}^{d-1} \lambda_{i}\left|\psi_{i}\right\rangle\left\langle\psi_{i}\right|$ can be written as\textsuperscript{\cite{Liu2019}}
\begin{small}
\begin{eqnarray}
\label{eq2}
\begin{split}
\mathcal{F}_{ab}=&\sum_{i=0}^{d-1} \frac{\left(\partial_{a} \lambda_{i}\right)\left(\partial_{b} \lambda_{i}\right)}{\lambda_{i}}\\
+&\sum_{i \neq j, \lambda_{i}+\lambda_{j} \neq 0} \frac{2\left(\lambda_{i}-\lambda_{j}\right)^{2}}{\lambda_{i}+\lambda_{j}} \operatorname{Re}\left(\left\langle\psi_{i}\mid\partial_{a}\psi_{j}\right\rangle\left\langle\partial_{b}\psi_{j}\mid\psi_{i}\right\rangle\right)
\end{split}
\end{eqnarray}
\end{small}

Notice that the QFI of parameter $\theta_{a}$ is just the diagonal element of the QFIM. In the independent estimation scenario, lower bound on the total variance of all parameters to be estimated in the system can be calculated by summing the inverse of diagonal
elements of the QFIM\textsuperscript{\cite{Braunstein1994}}
\begin{eqnarray}
\label{eq3}
\begin{split}
\delta^{\rm ind}=\sum_{k=1}^{n}\mathcal{F}_{kk}^{-1}
\end{split}
\end{eqnarray}

However, if we want to estimate the parameters simultaneously, a lower bound on the total variance of all parameters is determined by the trace of the inverse of the QFIM\textsuperscript{\cite{Humphreys2013}}
\begin{eqnarray}
\label{eq4}
\begin{split}
\delta^{\rm sim}=\operatorname{Tr}\mathcal{F}^{-1}(\boldsymbol{\theta})
\end{split}
\end{eqnarray}

To compare the performance of separate and simultaneous estimates, a ratio is defined as\textsuperscript{\cite{Yousefjani2017}}
\begin{eqnarray}
\label{eq5}
\begin{split}
R=\frac{\delta^{\rm ind}}{\frac{1}{n} \delta^{\rm sim}}
\end{split}
\end{eqnarray}
in which $n$ denotes the number of parameters to be estimated. The value of $R$ satisfies $0<R \leq n$ and $\mathrm{R}>1$ indicates that the performance of simultaneous estimation is better than that of individual estimation.

\subsection{Amplitude damping noise for qutrit}
The amplitude damping noise is the standard one for a dissipative interaction between a quantum system
and its environment.\textsuperscript{\cite{Nielsen2000}} For example, the model of AD noise can be applied to describe the photon loss of an optical field, or spontaneous emission of an atomic system weakly coupled to a zero temperature environment in the Born-Markov approximation.
For the qutrit-based system, the situation is more complicated since there are three different configurations to be considered, i.e., $V$, $\Lambda$ and $\Xi$ configurations.\textsuperscript{\cite{Hioe1981}} Here, we will focus on the $V$-configuration. If the environment is in a vacuum state, the amplitude damping noise of the $V$-type qutrit can be represented by the following Kraus operators\textsuperscript{\cite{Ch2007}}
\begin{eqnarray}
\label{eq6}
\begin{split}
E_{0}=&\left(\begin{array}{ccc}
1 & 0 & 0 \\
0 & \sqrt{1-d_1} & 0 \\
0 & 0 & \sqrt{1-d_2}
\end{array}\right),E_{1}=\left(\begin{array}{ccc}
0 & \sqrt{d_1} & 0 \\
0 & 0 & 0 \\
0 & 0 & 0
\end{array}\right)\\
E_{2}=&\left(\begin{array}{lll}
0 & 0 & \sqrt{d_2} \\
0 & 0 & 0 \\
0 & 0 & 0
\end{array}\right)
\end{split}
\end{eqnarray}
where $d_{1}=1-e^{-\gamma_{1} t}$ and $d_{2}=1-e^{-\gamma_{2} t}$. $\gamma_{1}$ and $\gamma_{2}$ are the spontaneous emission rates of the upper levels $|1\rangle$ and $|2\rangle$, respectively.

\subsection{Weak measurement and Quantum measurement reversal for qutrit}
The WM considered in this paper is the positive operator-valued measure (POVM) or partial-collapse measurement originally discussed 
by Korotkov.\textsuperscript{\cite{Korotkov1999,Korotkov2006}} In the qutrit case, the measurement operators can be expressed as\textsuperscript{\cite{Xiao2013}}
\begin{eqnarray}
\label{eq7}
\begin{split}
M_{0}=&\left(\begin{array}{ccc}
1 & 0 & 0 \\
0 & \sqrt{1-p} & 0 \\
0 & 0 & \sqrt{1-q}
\end{array}\right),
M_{1}=\left(\begin{array}{ccc}
0 & 0 & 0 \\
0 & \sqrt{p} & 0 \\
0 & 0 & 0
\end{array}\right)\\
M_{2}=&\left(\begin{array}{ccc}
0 & 0 & 0 \\
0 & 0 & 0 \\
0 & 0 & \sqrt{q}
\end{array}\right)
\end{split}
\end{eqnarray}
where $0 \leqslant p,q \leqslant 1$ describe the strengths of WM. Notice that the measurement operators $M_{1}$ and $M_{2}$ are equivalent to the von Neumann projective measurements which are irrevocable. Only the measurement operator $M_{0}$ is a WM for the qutrit system. Such a WM could be reversed by the operation of QMR which can be written as\textsuperscript{\cite{Xiao2013}}
\begin{eqnarray}
\label{eq8}
\begin{split}
M_{r}=\left(\begin{array}{ccc}
\sqrt{\left(1-p_{r}\right)\left(1-q_{r}\right)} & 0 & 0 \\
0 & \sqrt{1-q_{r}} & 0 \\
0 & 0 & \sqrt{1-p_{r}}
\end{array}\right)
\end{split}
\end{eqnarray}
where $0 \leqslant p_{r},q_{r} \leqslant 1$ are the strengths of QMR.

\subsection{Quantum Gates for qutrit}
The quantum circuit of qutrit teleportation involves some qutrit-based logic gates and operations. Here, we introduce the elementary quantum gates or operators for 3-dimension (3D) systems. The Not gate and phase gate for single qutrit are\textsuperscript{\cite{Fujii2001,Kari2002}}
\begin{eqnarray}
\label{eq9}
\begin{split}
X^{i}|m\rangle=|m \oplus i\rangle \\
Z^{k}|m\rangle=\omega^{k m}|m\rangle
\end{split}
\end{eqnarray}
where $\oplus$ denoting addition module $3$ and $\omega \equiv \exp (2 \pi i / 3)$.

The generalized Hadamard gate of qutrit is
\begin{small}
\begin{eqnarray}
\label{eq11}
\begin{split}
H=\frac{1}{\sqrt{3}} \sum_{m, n=0}^{2} \mathrm{e}^{2 \pi \mathrm{i} m n / 3}|m\rangle\langle n|=\frac{1}{\sqrt{3}}\left[\begin{array}{ccc}
1 & 1 & 1 \\
1 & \mathrm{e}^{2 \pi \mathrm{i} / 3} & \mathrm{e}^{4 \pi \mathrm{i} / 3} \\
1 & \mathrm{e}^{4 \pi \mathrm{i} / 3} & \mathrm{e}^{2 \pi \mathrm{i} / 3}
\end{array}\right]
\end{split}
\end{eqnarray}
\end{small}

$R_{C}$ and $L_{C}$ denotes the generalized CNOT Right-Shift gate and CNOT Left-Shift gate for two qutrits. They are as follows
\begin{eqnarray}
\label{eq10}
\begin{split}
R_{C}|m\rangle \otimes|n\rangle=|m\rangle \otimes|n \oplus m\rangle \\
L_{C}|m\rangle \otimes|n\rangle=|m\rangle \otimes|n \ominus m\rangle
\end{split}
\end{eqnarray}
with $\ominus$ denoting subtraction module $3$.

\section{Precision of phase estimation of the teleportated qutrit}
\label{sec3}
\begin{figure*}
\begin{center}
  \includegraphics[width=1\textwidth]{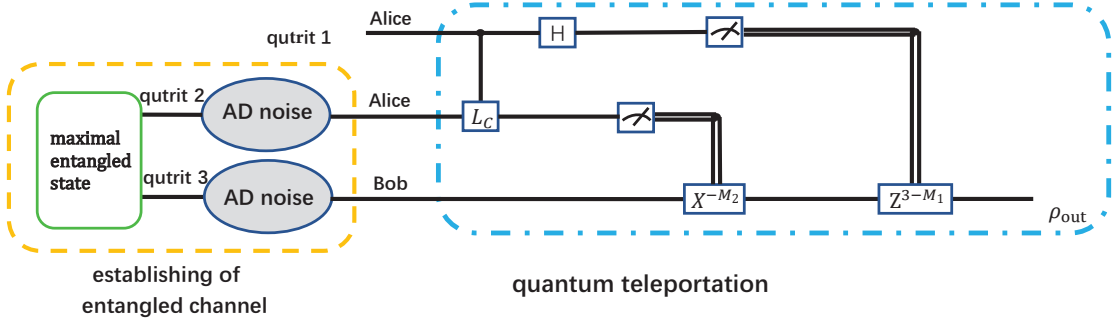}
\caption{(color online) Schematic illustrations of qutrit-based teleportation under the AD noise. }
\label{fig1}       
\end{center}
\end{figure*}

The two phase parameters $\phi_{1}$ and $\phi_{2}$ that we try to estimate are encoded in a qutrit state $|\psi\rangle_{\rm in}$
\begin{eqnarray}
\label{eq12}
\begin{split}
|\psi\rangle_{\rm in}= \alpha|0\rangle+ \beta e^{i \phi_{1}}|1\rangle+ \delta e^{i \phi_{2}}|2\rangle
\end{split}
\end{eqnarray}
where $\alpha$, $\beta$, $\delta$ are all real numbers and $\alpha^{2}+\beta^{2}+\delta^{2}=1$. Alice wishes to transfer the parameterized phase information to Bob through the procedure of teleportation. For this, they should establish the shared entanglement. Suppose Charlie has prepared a maximally entangled state in the form of
\begin{eqnarray}
\label{eq13}
\begin{split}
|\psi\rangle_{23}=\frac{1}{\sqrt{3}}(|00\rangle+|11\rangle+|22\rangle)
\end{split}
\end{eqnarray}
Then he distributes qutrits $2$ and $3$ to Alice and Bob through the independent AD channels, as shown in Figure \ref{fig1}. This represents a typical situation, where the shared entangled state is generated and distributed by the third party through the AD channel. Then the initially entangled state shown in Equation (\ref{eq13}) evolves into a mixed state in the presence of AD noise.
\begin{eqnarray}
\label{eq14}
\begin{split}
\boldsymbol{\rho}_{1}=\sum_{i,j} E_{i j} \boldsymbol{\rho}_{0} E_{i j}^{\dagger}
\end{split}
\end{eqnarray}
where $\boldsymbol{\rho}_{0}=\left|\psi\right\rangle_{23}\left\langle\psi\right|$, $E_{i j}=E_{i}^{(2)} \otimes E_{j}^{(3)}$, $i, j=0,1,2$. For the sake of simplicity, we will present the analytic expressions only for $d_{1}=d_{2}=d=1-e^{-\gamma t}$ because the general analytic expressions are too complicated to present.
After experiencing the AD noise, the non-zero elements of $\boldsymbol{\rho}_{1}$ the shared state in the basis of $\{|3j+k+1\rangle=|j,k\rangle\}$ can be written as the following 
\begin{eqnarray}
\label{eq15}
\begin{split}
\begin{aligned}
\rho_{11} &=\frac{1}{3}+\frac{2 d^{2}}{3} \\
\rho_{22} &=\rho_{33}=\rho_{44}=\rho_{77}=\frac{1}{3} d(1-d) \\
\rho_{55} &=\rho_{99}=\rho_{59}=\rho_{95}^{*}=\frac{1}{3}(1-d)^{2} \\
\rho_{15} &=\rho_{51}^{*}=\rho_{19}=\rho_{91}^{*}=\frac{1}{3}(1-d)
\end{aligned}
\end{split}
\end{eqnarray}

According to the circuit of quantum teleportation shown in \textbf{Figure~\ref{fig1}}, Bob finally obtains the output state $\boldsymbol{\rho}_{\text {out }}$ which can be given by
\begin{footnotesize}
\begin{eqnarray}
\label{eq16}
\begin{split}
\boldsymbol{\rho}_{\text {out }}=\left(\begin{array}{ccc}
\frac{A}{3}+(1-A) \alpha^{2} & B \alpha \beta e^{-i \phi_{1}} & B \alpha \delta e^{-i \phi_{2}} \\
B \alpha \beta e^{i \phi_{1}} & \frac{A}{3}+(1-A) \beta^{2} & B \beta \delta e^{-i\left(\phi_{2}-\phi_{1}\right)} \\
B \alpha \delta e^{i \phi_{2}} & B \beta \delta e^{i\left(\phi_{2}-\phi_{1}\right)} & \frac{A}{3}+(1-A) \delta^{2}
\end{array}\right)
\end{split}
\end{eqnarray}
\end{footnotesize}
where $A=2 d-2 d^{2}$ and $B=1+\frac{1}{3} d^{2}-\frac{4}{3} d$. In order to calculate the QFIM of the $\phi_{1}$ and $\phi_{2}$, one can diagonalize Equation~(\ref{eq16}) and obtain the non-zero eigenvalues and eigenvectors. Choosing $\alpha=\beta=\delta=\frac{1}{\sqrt{3}}$ and substituting them into Equation~(\ref{eq2}), the QFIM for the parameters $\phi_{1}$ and $\phi_{2}$ can be calculated as
\begin{eqnarray}
\label{eq17}
\begin{split}
\mathcal{F}_{\rm AD}=\left(\begin{array}{cc}
\frac{4 \sqrt{2}}{3} \frac{1}{\zeta_1} & \frac{4}{9} \frac{1}{\zeta_1} \\
\frac{4}{9} \frac{1}{\zeta_1} & \frac{4 \sqrt{2}}{3} \frac{1}{\zeta_1}
\end{array}\right)
\end{split}
\end{eqnarray}
where $\zeta_1=\frac{d^2-4 d+9}{\left(d^2-4 d+3\right)^2}$.

\begin{figure}
\begin{center}
  \includegraphics[width=0.5\textwidth]{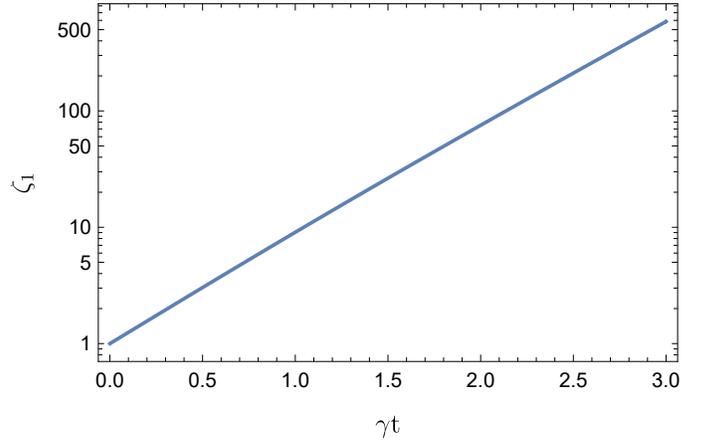}
\caption{(color online) $\zeta_1$ as a function of $\gamma t$. }
\label{fig2}       
\end{center}
\end{figure}

Then the total variances of phase parameters $\phi_{1}$ and $\phi_{2}$ for independent and simultaneous estimations yield to
\begin{eqnarray}
\label{eq18}
\begin{split}
\begin{aligned}
&\delta^{\rm ind}_{\rm AD}=\frac{3 \sqrt{2}}{4} \zeta_1 \\
&\delta^{\rm sim}_{\rm AD}=\frac{27 \sqrt{2}}{34} \zeta_1
\end{aligned}
\end{split}
\end{eqnarray}
It can be found that they are only determined by $\zeta_1$ but with different scaling factors.
\textbf{Figure~\ref{fig2}} shows $\zeta_1$ as a function of $\gamma t$. It is obvious that the value of $\zeta_1$ increases sharply with the increasing of $\gamma t$. The increasing of variance means that the estimation precision of the phase parameters $\phi_{1}$ and $\phi_{2}$ decrease rapidly as the AD noise increases. In the following, based on the techniques of WM and EAM, we propose two schemes to improve the precision of the two-parameter estimation under AD noise.

\section{Improving the precision of parameter estimation by WM and QMR}
\label{sec4}
In this section, we turn to show how the precision of parameter estimation can be improved by the synergistic action of WM and QMR. In the procedure of entanglement establishment, two WM operations are performed on the entangled qutrits before they pass through the AD noise, and finally two QMRs are carried out, respectively. After these operations, the entangled state of Equation~(\ref{eq13}) becomes
\begin{eqnarray}
\label{eq19}
\begin{split}
\boldsymbol{\rho}_{2}=\Big[\mathcal{M}_{r} \sum_{i j} E_{i j} (\mathcal{M}_{0} \boldsymbol{\rho}_{0} \mathcal{M}_{0}^{\dagger}) E_{i j}^{\dagger} \mathcal{M}_{r}^{\dagger}\Big]/W
\end{split}
\end{eqnarray}
where $\mathcal{M}_{0}=M_{0}^{(2)}\otimes M_{0}^{(3)}$ and $\mathcal{M}_{r}=M_{r}^{(2)}\otimes M_{r}^{(3)}$. $W=\frac{1}{3}\bar{p}_{r}^{2} \bar{q}_{r}^{2} \left( d^{2} \bar{p}^{2}+  d^{2} \bar{q}^{2}+ 1\right)+\frac{2}{3} d \bar{d} \bar{p}_{r} \bar{q}_{r} ( \bar{p}^{2} \bar{q}_{r} +\bar{p}_{r} \bar{q}^{2})+\frac{1}{3}\overline{d}^{2} (\bar{p}_{r}^{2} \bar{q}^{2}+ \bar{q}_{r}^{2} \bar{p}^{2})$ is the normalization parameter, and $\bar{o}= 1-o$. Assuming that qutrits $2$ and $3$ are identical and both subject to the same measurement strengths of WM and QMR. Then the non-zero elements of the shared entangled state $\boldsymbol{\rho}_{2}$ are as follows
\begin{eqnarray}
\label{eq20}
\begin{split}
\begin{aligned}
\rho_{11} &= \frac{1}{3W}\bar{p}_{r}^{2}\bar{q}_{r}^{2}[1+ d^{2}\bar{p}^{2}+ d^{2}\bar{q}^{2}] \\
\rho_{22} &=\rho_{44}=\frac{1}{3W} \bar{p}_{r}\bar{q}_{r}^{2}d\bar{d}\bar{p}^{2} \\
\rho_{33} &=\rho_{77}=\frac{1}{3W} \bar{p}_{r}^{2}\bar{q}_{r}d\bar{d}\bar{q}^{2} \\
\rho_{55} &=\frac{1}{3W}\bar{q}_{r}^{2}\overline{d}^{2}\bar{p}^{2} \\
\rho_{99} &=\frac{1}{3W}\bar{p}_{r}^{2}\overline{d}^{2}\bar{q}^{2} \\
\rho_{15} &=\rho_{51}^{*}=\frac{1}{3W}\bar{p}_{r}\bar{q}_{r}^{2}\bar{d}\bar{p} \\
\rho_{19} &=\rho_{91}^{*}=\frac{1}{3W}\bar{p}_{r}^{2}\bar{q}_{r}\bar{d}\bar{q} \\
\rho_{59} &=\rho_{95}^{*}=\frac{1}{3W}\bar{p}_{r}\bar{q}_{r}\overline{d}^{2}\bar{p}\bar{q}
\end{aligned}
\end{split}
\end{eqnarray}

Through the teleportation procedure shown in \textbf{Figure~\ref{fig1}}, the output state will be given as
\begin{footnotesize}
\begin{eqnarray}
\label{eq21}
\begin{split}
\boldsymbol{\rho}^{\rm WM}_{\text {out}}=\left(\begin{array}{ccc}
\frac{C}{3}+(1-C) \alpha^{2} & D \alpha \beta e^{-i \phi_{1}} & D \alpha \delta e^{-i \phi_{2}} \\
D \alpha \beta e^{i \phi_{1}} & \frac{C}{3}+(1-C) \beta^{2} & D \beta \delta e^{-i\left(\phi_{2}-\phi_{1}\right)} \\
D \alpha \delta e^{i \phi_{2}} & D \beta \delta e^{i\left(\phi_{2}-\phi_{1}\right)} & \frac{C}{3}+(1-C) \delta^{2}
\end{array}\right)
\end{split}
\end{eqnarray}
\end{footnotesize}
where $C= d \bar{d} \bar{p}_{r} \bar{q}_{r}( \bar{p}^{2}\bar{q}_{r} + \bar{q}^{2} \bar{p}_{r}) /W$ and $D= \frac{1}{3W} \bar{d} \bar{p}_{r} \bar{q}_{r}(\bar{p}_{r} \bar{q}+ \bar{d} \bar{p} \bar{q}+\bar{p}\bar{q}_{r})$.

Since we have previously assumed that $d_1=d_2=d$, it is consequently to assume $p=q$ and ${p}_{r}={q}_{r}$.
The QFIM for the parameters $\phi_{1}$ and $\phi_{2}$ can be obtained
\begin{eqnarray}
\label{eq22}
\begin{split}
&\mathcal{F}_{\rm WM}=\left(\begin{array}{cc}
\frac{4 \sqrt{2}}{3} \frac{1}{\zeta_2} & \frac{4}{9} \frac{1}{\zeta_2} \\
\frac{4}{9} \frac{1}{\zeta_2} & \frac{4 \sqrt{2}}{3} \frac{1}{\zeta_2}
\end{array}\right)
\end{split}
\end{eqnarray}
where $\zeta_2=\frac{f h+2 h^2}{3f^2}$, $f=\frac{1}{3} \bar{d} \bar{p} \bar{p}_{r}^{2}(2\bar{p}_{r}+\bar{d} \bar{p})$ and $h=\frac{1}{3}\bar{p}_{r}^{2}[\bar{p}_{r}^{2}(1+2 d^{2}\bar{p}^{2})+4 d \bar{d} \bar{p}^{2}\bar{p}_{r}+2 \overline{d}^{2} \bar{p}^{2}]$.

From Equations~(\ref{eq3}), (\ref{eq4}) and (\ref{eq22}), the lower bound for the independent and simultaneous estimations on the total variance of phase parameters $\phi_{1}$ and $\phi_{2}$ can be calculated as
\begin{eqnarray}
\label{eq23}
\begin{split}
\begin{aligned}
&\delta^{\rm ind}_{\rm WM}=\frac{3 \sqrt{2}}{4} \zeta_2 \\
&\delta^{\rm sim}_{\rm WM}=\frac{27 \sqrt{2}}{34} \zeta_2
\end{aligned}
\end{split}
\end{eqnarray}

As we all known, the smaller the variance, the higher the precision of multiparameter estimation. Therefore we want to minimize $\zeta_2$.
The minimum $\zeta_2$ can be derived from the conditions $\frac{\partial \zeta_2}{\partial p_{r}}=0$ and $\frac{\partial^{2} \zeta_2}{\partial p_{r}^{2}}>0$. The result turns out to be
\begin{eqnarray}
\label{eq24}
\begin{split}
\begin{aligned}
p_{r,\rm WM}^{\rm opt}=&\frac{1}{2(1+2 d^{2} \bar{p}^{2})}\Big[2+\bar{d}+d \bar{p}+2 d^{2}\bar{p}^{2}(2+\bar{p}\bar{d})\\
-&\bar{d}\bar{p} \sqrt{9-4 d \bar{p} (1-d\bar{p})^{2}(2-d\bar{p})} \Big]
\end{aligned}
\end{split}
\end{eqnarray}

\begin{figure*}
\begin{center}
  \includegraphics[width=1\textwidth]{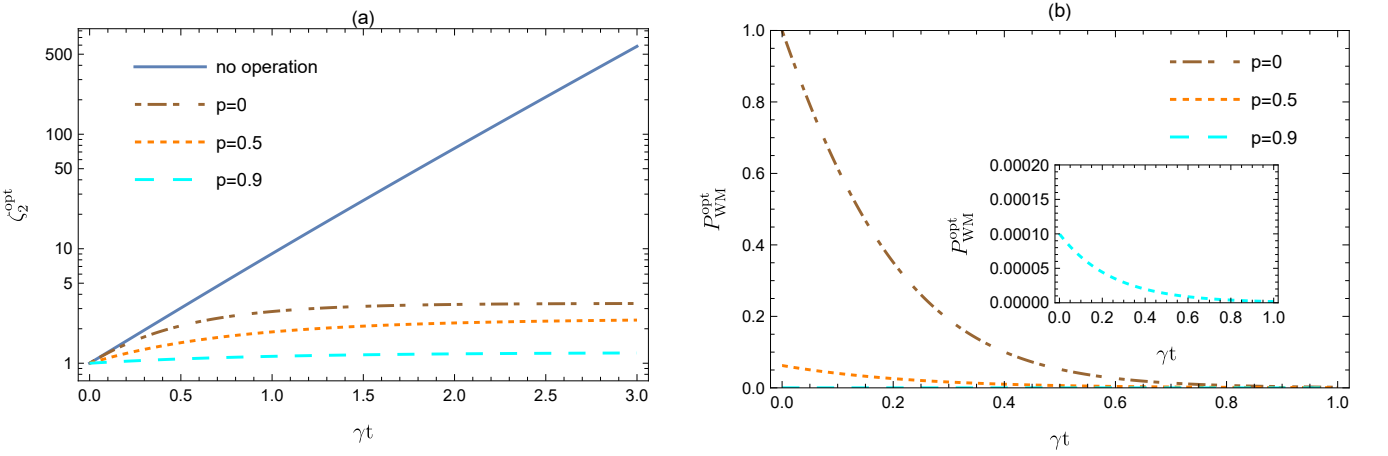}
\caption{(color online) (a) $\zeta_2^{\rm opt}$ as a function of $\gamma t$. (b) $P_{\rm WM}^{\rm opt}$ as a function of $\gamma t$ for different values of $p$.}
\label{fig3}       
\end{center}
\end{figure*}

Substituting the optimal QMR strength into Equation (\ref{eq23}), the optimal $\zeta_2^{\rm opt}$ can be easily obtained. Here, we only show the numerical results, because the general analytic expression is too complicated to present.
In \textbf{Figure~\ref{fig3}a} we plot the parameter $\zeta_2^{\rm opt}$ as a function of $\gamma t$ for different WM strengths $p$ under the optimal QMR strength. Unlike the results in \textbf{Figure~\ref{fig2}} where the total variances increases rapidly due to the AD noise (i.e., diverging in the no-operation case), \textbf{Figure~\ref{fig3}a} shows that WM and QMR can be indeed used for suppressing the total variances of both independent and simultaneous estimations and hence improving the estimation performance. In particular, in the case of $\gamma t>0.5$, the combined action of WM and QMR operations keeps the growth of $\zeta_2$ within a small range (i.e, converging in the WM cases). This means that this scheme works much better in the severe decoherence regime. Meanwhile, according to the results of different WM strengths in \textbf{Figure~\ref{fig3}a}, it is obvious that the larger the WM strength, the higher the estimation precision of the phase parameters $\phi_{1}$ and $\phi_{2}$. When the WM strength $p \rightarrow 1$, the estimation precision will not be affected by the AD noise.

The physical mechanism can be understood by two steps: Firstly, the pre-WM is performed to reduce the ratio of states $|1\rangle$ and $|2\rangle$ and increase the ratio of $|0\rangle$ of the qutrit. Since $|0\rangle$ is immune to AD noise, the state after WM becomes more robust to AD noise. Secondly, WM is a non-completely destructive measurement, the state after WM can be reversed by QMR in a probatilistic way. Therefore, the larger the WM strength $p$ is, the better the improvement of estimation precision under AD noise, as shown in \textbf{Figure~\ref{fig3}a}.

Due to the fact that both WM and QMR are non-unitary operations, such a method is not deterministic. The successful probability could be calculated by following the standard postulate of quantum measurement with two steps. Firstly, the probability of obtaining the measurement outcome of $\mathcal{M}_{0}=M_{0}^{(2)}\otimes M_{0}^{(3)}$ is $\mathcal{P}_{1}=\text{tr}\left(\mathcal{M}_{0}^{\dagger}\mathcal{M}_{0}\rho\right)$. Then, the reduced state will pass through the AD channel. Since the AD channel is trace-preserving, it doesn't change the successful probability. Secondly, the probability of obtaining the measurement outcome of QMR is $\mathcal{P}_{2}=\text{tr}\Big[ \mathcal{W}_{\rm tot}^{\dagger}\mathcal{W}_{\rm tot} \left(\sum_{i j} E_{i j} (\mathcal{M}_{0} \boldsymbol{\rho}_{0} \mathcal{M}_{0}^{\dagger}) E_{i j}^{\dagger}\right)\Big]$.
The final probability of this scheme depends on WM and QMR performed successfully in sequence, i.e., $P_{\rm WM}=\mathcal{P}_{1}\mathcal{P}_{2}$, which exactly equals to the normalization factor $W$. Under the condition of Equation (\ref{eq24}), the successful probability yields to
\begin{small}
\begin{eqnarray}
\label{eq25}
\begin{split}
\begin{aligned}
P_{\rm WM}^{\rm opt}=&(1-p_{r,\rm WM}^{\rm opt})^4 \left(\frac{2}{3} d^2 \bar{p}^2+\frac{1}{3}\right)+\frac{4}{3} (1-p_{r,\rm WM}^{\rm opt})^3 d \bar{d} \bar{p}^2\\
+&\frac{2}{3} (1-p_{r,\rm WM}^{\rm opt})^2 \overline{d}^2 \bar{p}^2
\end{aligned}
\end{split}
\end{eqnarray}
\end{small}
It can be seen that the successful probability decreases with the increasing strength of WM, as shown in \textbf{Figure~\ref{fig3}b}. This result implies that the improvement of the estimation precision is at the price of reducing the probability of success.

\section{Improving the precision of parameter estimation by EAM and QMR}
\label{sec5}
In this section, we present another scheme to improve the precision of parameter estimation under the AD noise by EAM and QMR. The idea is based on the fact that some of the Kraus operators $E_{ij}$ of AD noise maybe reversible. From Equation~(\ref{eq6}), we will find that only $E_{00}$ is reversible. Thus, if we can choose the reversible Kraus operator $E_{00}$ during the evolution, then we can reverse the impact of AD noise by appropriate QMR. The selection of
$E_{00}$ is implemented by monitoring the environment coupled with the system. Only when the detection of environment is null, then the following QMR operation is carried out. The other measurement results with clicks are discarded.

Initially, two qutrits are prepared in the entangled state shown in Equation~(\ref{eq13}) and both the environments are in vacuum state. Suppose the outcome of EAM is no click, then the entangled state will be mapped to
\begin{eqnarray}
\label{eq26}
\begin{split}
\boldsymbol{\rho}_{3}=\Big(\mathcal{M}_{r} E_{0 0} \boldsymbol{\rho}_{0} E_{0 0}^{\dagger} \mathcal{M}_{r}^{\dagger}\Big)/U
\end{split}
\end{eqnarray}
where $U=\frac{1}{3} (\bar{p}_{r}^2 \bar{q}_{r}^2+ \overline{d}^2 \bar{q}_{r}^2+ \overline{d}^2 \bar{p}_{r}^2)$ is the normalization parameter. The non-zero elements of $\boldsymbol{\rho}_{3}$ are
\begin{eqnarray}
\label{eq27}
\begin{split}
\begin{aligned}
\rho_{11} &=\frac{1}{3U} \bar{p}_{r}^2 \bar{q}_{r}^2 \\
\rho_{55} &=\frac{1}{3U} \overline{d}^2 \bar{q}_{r}^2 \\
\rho_{99} &=\frac{1}{3U} \overline{d}^2 \bar{p}_{r}^2 \\
\rho_{15} &=\rho_{51}^{*}=\frac{1}{3U} \bar{p}_{r} \bar{d} \bar{q}_{r}^2 \\
\rho_{19} &=\rho_{91}^{*}=\frac{1}{3U} \bar{p}_{r}^2 \bar{d} \bar{q}_{r} \\
\rho_{59} &=\rho_{95}^{*}=\frac{1}{3U} \bar{p}_{r} \overline{d}^2 \bar{q}_{r}
\end{aligned}
\end{split}
\end{eqnarray}

Following the quantum teleportation protocol shown in \textbf{Figure \ref{fig1}}, the output state can be obtained
\begin{small}
\begin{eqnarray}
\label{eq28}
\begin{split}
\boldsymbol{\rho}^{\text {EAM }}_{\rm out}=\left(\begin{array}{ccc}
\alpha^{2} & G \alpha \beta e^{-i \phi_{1}} & G \alpha \delta e^{-i \phi_{2}} \\
G \alpha \beta e^{i \phi_{1}} & \beta^{2} & G \beta \delta e^{-i\left(\phi_{2}-\phi_{1}\right)} \\
G \alpha \delta e^{i \phi_{2}} & G \beta \delta e^{i\left(\phi_{2}-\phi_{1}\right)} & \delta^{2}
\end{array}\right)
\end{split}
\end{eqnarray}
\end{small}
where $G=\frac{\bar{d}\bar{p}_{r}\bar{q}_{r}\left(\bar{d}+\bar{p}_{r}+\bar{q}_{r}\right)}{3U}$.

We still assume that ${p}_{r}={q}_{r}$, then the QFIM for the parameters $\phi_{1}$ and $\phi_{2}$ can be obtained by diagonalizing $\boldsymbol{\rho}_{\text {out }}^{\rm EAM}$, which gives
\begin{eqnarray}
\label{eq29}
\begin{split}
&\mathcal{F}_{\rm EAM}=\left(\begin{array}{cc}
\frac{4 \sqrt{2}}{3} \frac{1}{\zeta_3} & \frac{4}{9} \frac{1}{\zeta_3} \\
\frac{4}{9} \frac{1}{\zeta_3} & \frac{4 \sqrt{2}}{3} \frac{1}{\zeta_3}
\end{array}\right)
\end{split}
\end{eqnarray}
where $\zeta_3=\frac{uv+2 v^2}{3u^2}$, $u=\bar{d}(\bar{d}+2\bar{q}_{r}^{2})$ and $v=\bar{q}_{r}^{2}+2 \overline{d}^2$. 

\begin{figure*}
\begin{center}
  \includegraphics[width=1\textwidth]{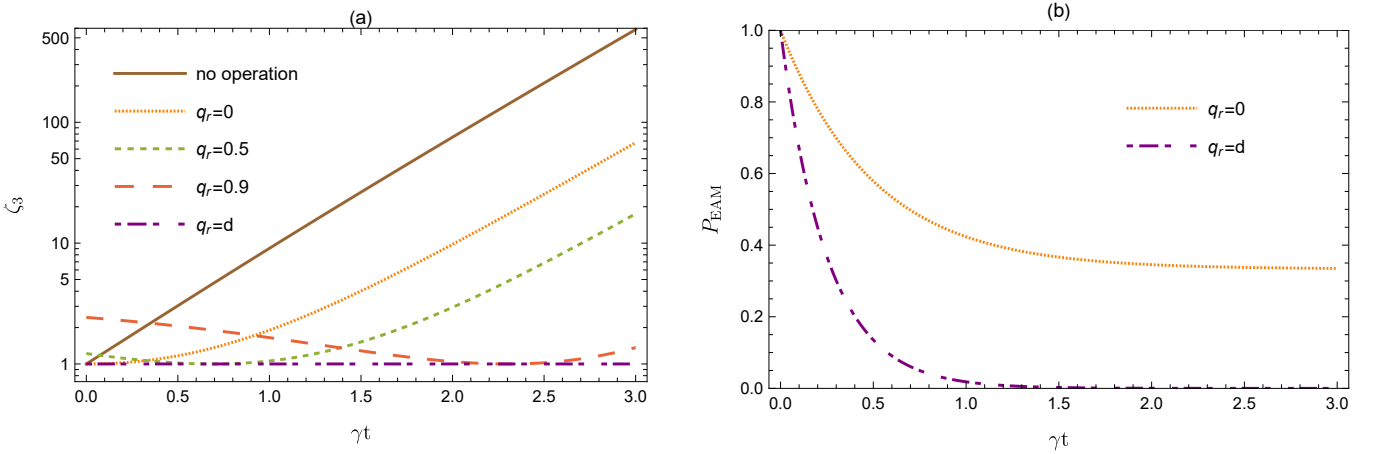}
\caption{(color online) (a) $\zeta_3$ as a function of $\gamma t$. (b) $P_{\rm EAM}$ as a function $\gamma t$ for different values of ${q}_{r}$}. 
\label{fig4}       
\end{center}
\end{figure*}

To demonstrate the power of EAM and QMR operations on the multiparameter estimation, we calculate the lower bound of the total variances of the phase parameters $\phi_{1}$ and $\phi_{2}$ for the independent and simultaneous estimations
\begin{eqnarray}
\label{eq30}
\begin{split}
\begin{aligned}
&\delta^{\rm ind}_{\rm EAM}=\frac{3 \sqrt{2}}{4} \zeta_3 \\
&\delta^{\rm sim}_{\rm EAM}=\frac{27 \sqrt{2}}{34} \zeta_3
\end{aligned}
\end{split}
\end{eqnarray}

Similarly, the optimal strength $q_{r}$ corresponding to the minimal $\zeta_3$ can be derived from the conditions $\frac{\partial \zeta_3}{\partial q_{r}}=0$ and $\frac{\partial^{2} \zeta_3}{\partial q_{r}^{2}}>0$. The result is given by
\begin{eqnarray}
\label{eq31}
\begin{split}
\begin{aligned}
q_{r,\rm EAM}^{\rm opt}=d
\end{aligned}
\end{split}
\end{eqnarray}

\textbf{Figure~\ref{fig4}a} shows the dynamics of parameter $\zeta_3$ as a function of $\gamma t$ for different QMR strengths $q_{r}$. It is obvious that the choose of QMR plays a significant role in this scheme. When the QMR is not performed (i.e., $q_{r}= 0$) or not selected to the optimal situation, $\zeta_3$ still diverges after a short time period. In the case of $q_{r}=0.5$, we find the lower bound of the total variance will first reach the minimal value and then begin to increase after passing through the minimum. As the QMR strength $q_{r}$ increases, the minimum of total variance will appear later but the trend of $\zeta_3$ seems to be diverged in the long time limit. However, by choosing the optimal strength of QMR, i.e., $q_{r,\rm EAM}^{\rm opt}=d$, it is intriguing to note that the lower bound of the total variance will always be kept at the optimal value, thus the improvement of parameter estimation precision will be the best at this time.
The underlying reason is that the introducing of EAM post-selects the result of quantum system suffered from the
decoherence process $E_{0 0}$. As we mentioned before, such process can be reversed by a proper QMR.

Considering that both EAM and QMR are non-unitary operations, the scheme is also probatilistic. The successful probability is given by
\begin{eqnarray}
\label{e36}
\begin{split}
\begin{aligned}
P_{\rm EAM}=\frac{1}{3} (1-q_{r})^4+\frac{2}{3} (1 - d)^2 (1 - q_{r})^2
\end{aligned}
\end{split}
\end{eqnarray}
which reduces to $P_{\rm EAM}^{\rm opt}=(1-d)^4$ when $q_{r,\rm EAM}^{\rm opt}=d$.
\textbf{Figure~\ref{fig4}b} sketches the success probability of the EAM scheme as a function of $\gamma t$. Here, we only choose two cases: without QMR $q_{r}=0$ and with the optimal QMR $q_{r}=d$, because the other cases are unhelpful.
The dot-dashed line denotes the success probability $P_{\rm EAM}^{\rm opt}$ with which the total variance achieves the minimal value. A simple comparison between \textbf{Figure~\ref{fig3}b} and \textbf{Figure~\ref{fig4}b} shows that the success probability of EAM scheme is higher than that of the WM scheme. Particularly, in the case of $\gamma t=0.5$, the success probability $P_{\rm WM}$ tends to be close to 0 while $P_{\rm EAM}^{\rm opt}$ is larger than 0.1.

\section{Discussions and Conclusions}
\label{sec6}
We have shown that both WM scheme and EAM scheme enable to improve the multiparameter estimation precision by selecting the appropriate measurement strength of QMR. However, the probabilistic nature of them inspires us to consider the efficiency of them in improving the precision of multiparameter estimation. To ensure an equitable comparison of these two schemes, we introduce the difference in an average improvement of the precision with both schemes work at the optimal measurement strength of QMR.
\begin{eqnarray}
\label{e37}
\begin{split}
\Delta=\frac{1}{\zeta_{3}^{\rm opt}} \times P_{\rm EAM}^{\rm opt}-\frac{1}{\zeta_{2}^{\rm opt}} \times P_{\rm WM}^{\rm opt}
\end{split}
\end{eqnarray}
The quantity $\Delta$ measures the superiority of EAM scheme compared to WM scheme in terms of improving the precision of multiparameter estimation. 
\begin{figure}
\begin{center}
  \includegraphics[width=0.5\textwidth]{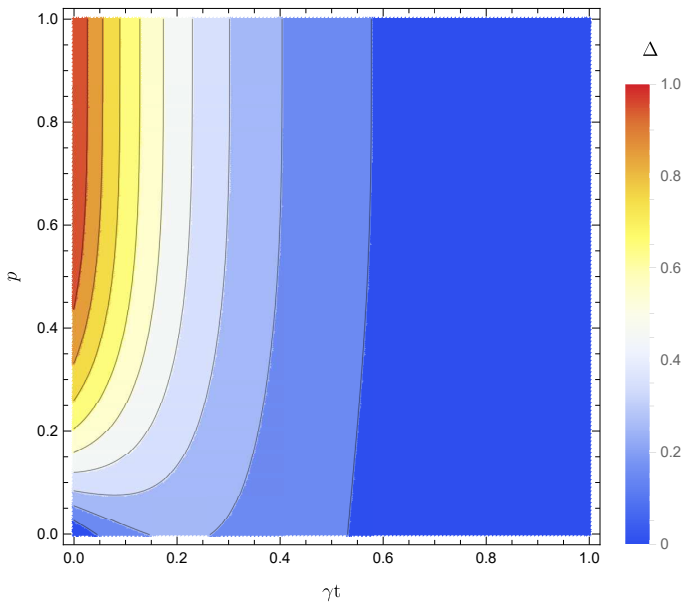}
\caption{(color online) The difference $\Delta$ versus $\gamma t$ and the WM strength $p$.}
\label{fig5}       
\end{center}
\end{figure}

The behavior of $\Delta$ as a function of $\gamma t$ and $p$ is shown in \textbf{Figure~\ref{fig5}}. An interesting finding is that $\Delta$ is always larger than 0. Especially in the regime of $\gamma t<0.5$, the efficiency of EAM scheme is significantly better than that of WM scheme. Such an advantage stems from the different order of action between the two types of measurement and AD noise. The WM is performed before the qutrit suffered AD noise, while the EAM is carried out after the AD noise. This means that the EAM not only collects the information from the system but also gathers the information from the environment. As a consequence, the EAM scheme more effectively improves the estimation precision in the qutrit teleportation.

Another interesting problem needs to be clarified is whether the superiority of simultaneous estimation still holds for these two schemes. For two-parameter estimation, we have $n=2$. According to the Equations (\ref{eq18}), (\ref{eq23}), and (\ref{eq30}), the ratio of Equation (\ref{eq5}) turns out to be $R=R_{\rm WM}=R_{\rm EAM}=\frac{17}{9}$. The ratio $R=\frac{17}{9}>1$ indicates that the performance of simultaneous estimation is always better than that of individual estimation regardless of whether WM and EAM are involved or not.

In summary, we revisited the multiparameter estimation problem from the perspective of quantum teleportation, focusing on two phase parameters $\phi_{1}$ and $\phi_{2}$ encoded in a teleported qutrit state. We proposed two schemes for combating the AD noise and improving the estimation precision. The first one is based on WM and QMR. We showed that the precisions of independent and simultaneous estimations can be improved with the same scaling factor. However, the efficiency is highly dependent on the strength of WM. Only when $p\rightarrow 1$, the best estimation precision can be achieved but with a very low success probability. The second scheme is based on EAM and QMR. Crucially, we found that the best estimation precision can be maintained by selecting the appropriate strength of QMR. We adopted a quantity $\Delta$ which takes into account both estimation precision and probability of success, as figure of merit to account for the performance of WM and EAM schemes. As one might expect, the EAM scheme works more precisely and efficiently because the post-measurement of EAM collects the information of system and environment at the same time. We also highlighted that simultaneous estimation is always advantageous regardless of whether WM and EAM are involved or not.
Our model realizes the teleportation of multiparameter information for a qutrit system and improves the estimation precision through two different schemes. Although both schemes are probabilistic, we believe that they still provide a new perspective of actively battling against decoherence in the scenarios remote sensing.

\begin{acknowledgements}
This work is supported by Jiangxi Provincial Natural Science Foundation under Grant Nos. 20212ACB211004 and 20212BAB201014. And it is supported by the Funds of the National Natural Science Foundation of China under Grant Nos. 12265004 and 61765007.
\end{acknowledgements}



\end{document}